\documentstyle[12pt,epsfig]{article}
\begin{document}
\setlength{\parskip}{0.45cm}
\setlength{\baselineskip}{0.75cm}
%
%
%
\begin{titlepage}
\setlength{\parskip}{0.25cm}
\setlength{\baselineskip}{0.25cm}
\begin{flushright}
DO-TH 2000/03\\  
\vspace{0.2cm}
January 2000
\end{flushright}
\vspace{1.0cm}
\begin{center}
\LARGE
{\bf Phenomenology of the Flavor--Asymmetry\\
\vspace{0.1cm}
 in the Light--Quark Sea of the Nucleon}
\vspace{1.5cm}

\large
M. Gl\"uck and E.\ Reya\\
\vspace{1.0cm}

\normalsize
{\it Universit\"{a}t Dortmund, Institut f\"{u}r Physik,}\\ 
{\it D-44221 Dortmund, Germany} \\

\vspace{1.5cm}
\end{center}
\begin{abstract}
A phenomenological ansatz for the flavor--asymmetry of the light sea
distributions of the nucleon, based on the Pauli exclusion principle,
is proposed.  This ansatz is compatible with the measured flavor--asymmetry
of the unpolarized sea distributions, $\bar{d}>\bar{u}$, of the nucleon.
A prediction for the corresponding polarized flavor--asymmetry is
presented and shown to agree with predictions of (chiral quark--soliton)
models which successfully reproduced the flavor--asymmetry of the 
unpolarized sea.
\end{abstract}
\end{titlepage}
%

The flavor--asymmetry of the light--quark sea in the nucleon has
attracted a lot of attention and many attempts were undertaken to 
explain the origin and calculate its magnitude (e.g., Ref.\ \cite{ref1}
and references therein).  In the present article we study this issue
inspired by the suggestion \cite{ref2} that this asymmetry is related
to the Pauli exclusion principle (`Pauli blocking').

Our proposed implementation of this idea is summarized by the 
phenomenological ansatz for the unpolarized and polarized antiquark
distributions
\begin{equation}
\bar{d}(x,Q_0^2)/\bar{u}(x,Q_0^2) = u(x,Q_0^2)/d(x,Q_0^2)
\end{equation}
\vspace{-1.5cm}

\noindent and

\vspace{-1.2cm}

\begin{equation}
\Delta\bar{d}(x,Q_0^2)/\Delta\bar{u}(x,Q_0^2) =
    \Delta u(x,Q_0^2)/\Delta d(x,Q_0^2)\, ,
\end{equation}
\vspace{-1.0cm}

\noindent respectively, with $Q_0^2$ being some low resolution scale, e.g., 
the one in \cite{ref3,ref4}.  These are our basic relations for the 
flavor--asymmetries of the unpolarized and polarized light sea densities
which imply that $u>d$ determines $\bar{u}<\bar{d}$, etc.  This is in
accordance with the suggestion of Feynman and Field \cite{ref2} that,
since there are more $u$- than $d$--quarks in the proton, $u\bar{u}$
pairs in the sea are suppressed more than $d\bar{d}$ pairs by the
exclusion principle.  It should be emphasized that our suggested regularity
in (1) is entirely of {\underline{empirical}} origin, and our anticipated
relation (2) has of course to be tested by future polarized experiments.
Both relations require obviously the idealized situation of maximal
Pauli--blocking and hold approximately in some effective field theoretic
(mesonic, bag and chiral) models \cite{ref1} as we shall see below. 

In Table I we present $\bar{d}(x,Q_0^2)/\bar{u}(x,Q_0^2)$ calculated
according to Eq.\ (1) from the (fitted) $d,\, u$ input distributions
of GRV98 \cite{ref5}, as compared to the actual fitted values of this
ratio.  The good agreement lends support to the phenomenological ansatz
in Eq.\ (1) and thus also to the experimentally so far unknown polarized
antiquark flavor--asymmetry implied by Eq.\ (2).  The predictions for
$\Delta\bar{d}/\Delta\bar{u}$ according to Eq.\ (2) are shown in Table II
utilizing the most recent LO AAC \cite{ref6} distributions $\Delta u
(x,Q_0^2)$ and $\Delta d(x,Q_0^2)$ which compare favorably with the 
predictions of the relativistic field theoretical chiral quark--soliton 
model \cite{ref7} for $\Delta\bar{d}/\Delta\bar{u}$, as well as with a
recent analysis based on the statistical parton model \cite{ref8}. The latter 
flavor--asymmetry for $\Delta\bar{u}$ and $\Delta\bar{d}$ can also be studied by replacing
the common constraint $\Delta\bar{u}=\Delta\bar{d}\equiv \Delta\bar{q}$
by our present Eq.\ (2).  Using the recent analysis of \cite{ref6}, for
example, one obtains the LO results for $\Delta\bar{u}(x,Q_0^2)$,
$\Delta \bar{d}(x,Q_0^2)$ and their difference presented in Figs.\ 1
and 2, respectively \cite{ref9}.  These predictions refer to an input
scale of $Q_0^2=1$ GeV$^2$ \cite{ref6}.  At the somewhat lower dynamical
input scales $Q_0^2=0.3-0.4$ GeV$^2$ \cite{ref3,ref4,ref5}, the 
maxima/minima of the curves shown in Figs.\ 1 and 2 move slightly to
the right, i.e.\ to slightly larger values of $x$.  Strictly speaking
a more consistent study of the antiquark asymmetry should be done 
\cite{ref10} within the framework of the `valence' scenario \cite{ref4} 
where $\Delta s(x,Q_0^2) = \Delta\bar{s}(x,Q_0^2)=0$.  This,
however, is expected to modify the present results only marginally.
The NLO analysis of the polarized antiquark asymmetry \cite{ref10} 
affords a {\underline{direct}} implementation of Eq.\ (2) in the fit
procedure due to the enhanced sensitivity of the NLO calculation of
$g_1^{p,n}(x,Q^2)$  to the polarized gluon distribution which is
affected by modifications of the polarized quark and antiquark 
distributions.  Again, no qualitative changes of our present results
are expected.

It is interesting to note that our results for the flavor--asymmetry
of the polarized sea distributions at $Q_0^2=1$ GeV$^2$ in Figs.\ 1
and 2 are comparable to those obtained in chiral quark--soliton model
calculations \cite{ref7} and in the previously mentioned statistical
parton model \cite{ref8} which correctly reproduced
the flavor--asymmetry of the unpolarized sea \cite{ref11,ref8}! 

A further {\underline{direct}} test of our phenomenological ansatz (2)
must await the polarized version \cite{ref12} of the Drell--Yan  
$\mu^+\mu^-$
pair production experiments \cite{ref13} which provided the information
on the flavor--asymmetry of the unpolarized sea distributions  
\mbox{in Eq.\ (1).}
\mbox{Future polarized} semi--inclusive DIS experiments at CERN (COMPASS) 
and \mbox{DESY} (HERMES) could also become relevant for measuring possible 
flavor--asymmetries of polarized light--quark sea distributions 
\cite{ref14}.  The statistics of present SMC \cite{ref15} and
HERMES \cite{ref16} measurements is not sufficient for testing our
expectations in \mbox{Figs.\ 1 and 2,} for \mbox{example}, despite 
the fact that
rather stringent model assumptions have been made for the analyses
of these experiments in order to improve the statistical significance.

Finally it should be noted that the data select the solution of Eq.\ (2)
which satisfies
\begin{equation}
\Delta q(x,Q_0^2)\, \Delta\bar{q}(x,Q_0^2)>0
\end{equation}
where $q=u,\, d$.  This can be understood as a consequence of the
expected predominant {\underline{pseudoscalar}} configuration of
the quark--antiquark pairs in the nucleon sea.  In fact, Eqs.\ (1) 
and (2) can be rewritten as
\begin{eqnarray}
u_+\bar{u}_+ + u_-\bar{u}_- & = & d_+\bar{d}_+ +d_- \bar{d}_- \equiv f_p\\
u_+\bar{u}_- + u_-\bar{u}_+ & = & d_+\bar{d}_- +d_- \bar{d}_+ \equiv f_a
\end{eqnarray}
where the common helicity densities are given by 
$\stackrel{(-)}{q}\!\!_{\pm}  
=(\stackrel{(-)}{q}\pm\,\,\Delta\!\!\!\stackrel{(-)}{q})/2$
and, for brevity, we dropped the $x$--dependence everywhere.
A predominant (pseudo)scalar configuration of the $(q\bar{q})$ pairs
in the nucleon sea implies, via Pauli--blocking, that the aligned 
quark--quark configurations 
$q_+(q_+\bar{q}_-)$ and $q_-(q_-\bar{q}_+)$ 
are suppressed relatively to the antialigned $q_+(q_-\bar{q}_+)$
and $q_-(q_+\bar{q}_-)$ `cloud' configurations, i.e.\ $f_p>f_a$ which 
yields
the result in  \mbox{Eq.\ (3)}.
\vspace{0.8cm}


We would like to thank M.\ Stratmann and W.\ Vogelsang for helpful
discussions and comments.
This work has been supported in part by the `Bundesministerium 
f\"ur Bildung, Wissenschaft, Forschung und Technologie', Bonn.

\newpage

\noindent{\large{\bf{\underline{Table I.}}}}\hspace{0.25cm}  The predicted
flavor--asymmetry $\bar{d}(x,Q_0^2)/\bar{u}(x,Q_0^2)$ according to Eq.\ (1)
using the GRV98 \cite{ref5} distributions $u(x,Q_0^2)$ and $d(x,Q_0^2)$
at the LO-- and NLO--QCD input scales $Q_0^2\equiv\mu_{\rm{LO}}^2\,\,
(\mu_{\rm{NLO}}^2) = 0.26$ GeV$^2\,\,(0.4$ GeV$^2)$.  
The actually fitted \cite{ref5}
ratio $(\bar{d}/\bar{u})_{\rm{fit}}$ is shown as well for comparison.
The NLO results are shown in parentheses.
\vspace{0.15cm}
\begin{center}
\begin{tabular}{c||c|c|c|c|c}
\renewcommand{\arraystretch}{2mm}
x & 0.01 & 0.05 & 0.1 & 0.2 & 0.3\\[1mm]
\hline\hline 
& & & & & \\
$\bar{d}/\bar{u}$ & 1.12 (1.13) & 1.25 (1.24) 
    & 1.34 (1.36) & 1.64 (1.70) & 2.04 (2.11)\\[4mm] 
$(\bar{d}/\bar{u})_{\rm{fit}}$ & 1.03 (1.04) & 1.16 (1.19) 
         & 1.50 (1.53) & 2.00 (1.84) & 1.98 (1.65)\\
\end{tabular}
\end{center}
\vspace{1.0cm}

\noindent{\large{\bf{\underline{Table II.}}}}\hspace{0.25cm}  The predicted
polarized flavor--asymmetry $\Delta\bar{d}(x,Q_0^2)/\Delta\bar{u}(x,Q_0^2)$ 
\mbox{according} to Eq.\ (2) using the LO AAC input distributions $\Delta 
u(x,Q_0^2)$ and $\Delta d(x,Q_0^2)$ at $Q_0^2=1$ GeV$^2$ \cite{ref6}.
The NLO results are similar.  The chiral quark--soliton predictions for
$\Delta\bar{d}/\Delta\bar{u}$ of Wakamatsu and Kubota \cite{ref7} are shown
for comparison as well which refer to a scale $Q_0^2\simeq 0.36$ GeV$^2$.

\begin{center}
\begin{tabular}{cc||c|c|c|c|c}
 & x & 0.01 & 0.05 & 0.1 & 0.2 & 0.3\\[1mm]
\hline\hline
 & & & & & & \\
\raisebox{-0.4cm}{$\frac{-\Delta\bar{d}}{\Delta\bar{u}}$} & AAC 
       & 1.55 & 1.76  & 1.95 & 2.26 & 2.59\\[4mm] 
 & soliton & 1.73 & 2.05 & 2.15 & 1.94 & 1.66\\
\end{tabular}
\end{center}
\vspace{1.3cm}

\noindent{\large{\bf{\underline{Figure Captions}}}}
\begin{itemize}
\item[\bf{Fig.\ 1}]  The predictions for the polarized sea distributions
        $\Delta\bar{u}(x,Q_0^2)$ and $\Delta\bar{d}(x,Q_0^2)$ 
        \mbox{according}
        to Eq.\ (2) with the LO results for $\Delta u(x,Q_0^2)$ and
        $\Delta d(x,Q_0^2)$ taken from Ref.\ \cite{ref6} at $Q_0^2=1$
        GeV$^2$. 

\item[\bf{Fig.\ 2}]  The same as in Fig.\ 1 but for $x\Delta\bar{u}(x,Q_0^2)
        - x\Delta\bar{d}(x,Q_0^2)$ at $Q_0^2=1$ GeV$^2$.
\end{itemize}
\newpage
\pagestyle{empty}
\begin{figure}
\vspace*{-2.5cm}
\hspace*{-1.5cm}
\epsfig{figure=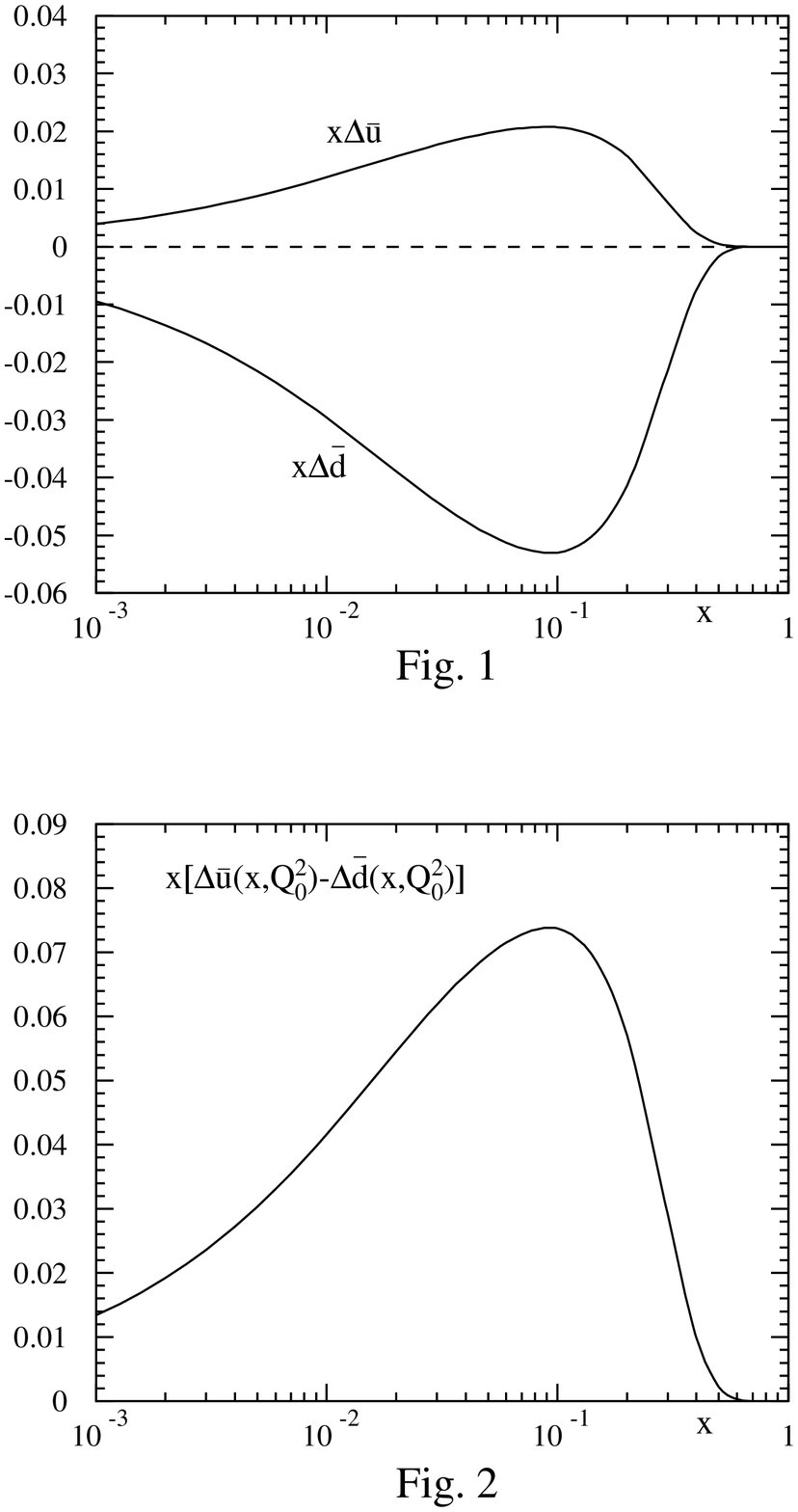,width=20cm}
\end{figure}
\end{document}